\documentclass[aps,prl,twocolumn,groupedaddress,showpacs]{revtex4}
\usepackage{graphicx}
\usepackage{amsmath}
\usepackage{mathrsfs}
\usepackage{amsmath}
\usepackage{amsfonts}
\usepackage{amssymb}

\begin{document}
\title{Entanglement of Atomic Qubits using an Optical Frequency Comb}
\author{D. Hayes}
\email[dhayes12@umd.edu]{}
\author{D. N. Matsukevich}
\author{P. Maunz}
\author{D. Hucul}
\author{Q. Quraishi}
\author{S. Olmschenk}
\author{W. Campbell}
\author{J. Mizrahi}
\author{C. Senko}
\author{C. Monroe}
\affiliation{Joint Quantum Institute and Department of Physics,
University of Maryland, College Park, MD 20742, USA}

\begin{abstract}
We demonstrate the use of an optical frequency comb to coherently control and entangle atomic qubits.  A train of off-resonant ultrafast laser pulses is used to efficiently and coherently transfer population between electronic and vibrational states of trapped atomic ions and implement an entangling quantum logic gate with high fidelity.  This technique can be extended to the high field regime where operations can be performed faster than the trap frequency. This general approach can be applied to more complex quantum systems, such as large collections of interacting atoms or molecules.
\end{abstract}
\date{\today}
\maketitle

The optical frequency comb generated from an ultrafast laser pulse train has revolutionized optical frequency metrology \cite{udem:2002,cundiff:2003,hall:2006,hansch:2006} and is now playing an important role in high resolution spectroscopy \cite{stowe:2008}.  The spectral purity yet large bandwidth of optical frequency combs also provides a means for the precise control of generic quantum systems, with examples such as the quantum control of multilevel atomic systems \cite{stowe:2006,stowe:PRL:2008}, laser cooling of molecules or exotic atomic species \cite{kielpinski:2006,viteau:2008}, and quantum state engineering of particular rovibronic states in molecules \cite{peer:2007,shapiro:2008}.  The optical frequency comb may become a crucial component in the field of quantum information science, where complex multilevel quantum systems must be controlled with great precision \cite{coherence:2008}.  

In this Letter, we report the use of an optical frequency comb generated from an ultrafast mode-locked laser to efficiently control and faithfully entangle two trapped atomic ion qubits.  The optical pulse train drives stimulated Raman transitions between hyperfine levels \cite{mlynek:1981,fukuda:1981}, accompanied by qubit state dependent momentum kicks \cite{poyatos:1996}.  The coherent accumulation of these pulses generates particular quantum gate operations that are controlled through the phase relationship between successive pulses.  This precise spectral control of the process along with the large optical bandwidth required for bridging the qubit frequency splitting forms a simple method for controlling both the internal electronic and external motional states of trapped ion qubits, and may be extended to most atomic species.  This same approach can be applied to control larger trapped ion crystals with more advanced pulse-shaping techniques, and can also be extended to a strong pulse regime where only a few high power pulses are needed for fast quantum gate operations in trapped ions \cite{poyatos:1996,garcia:2003,duan:2004}.

\begin{figure}[!]
\includegraphics[width=2.5in]{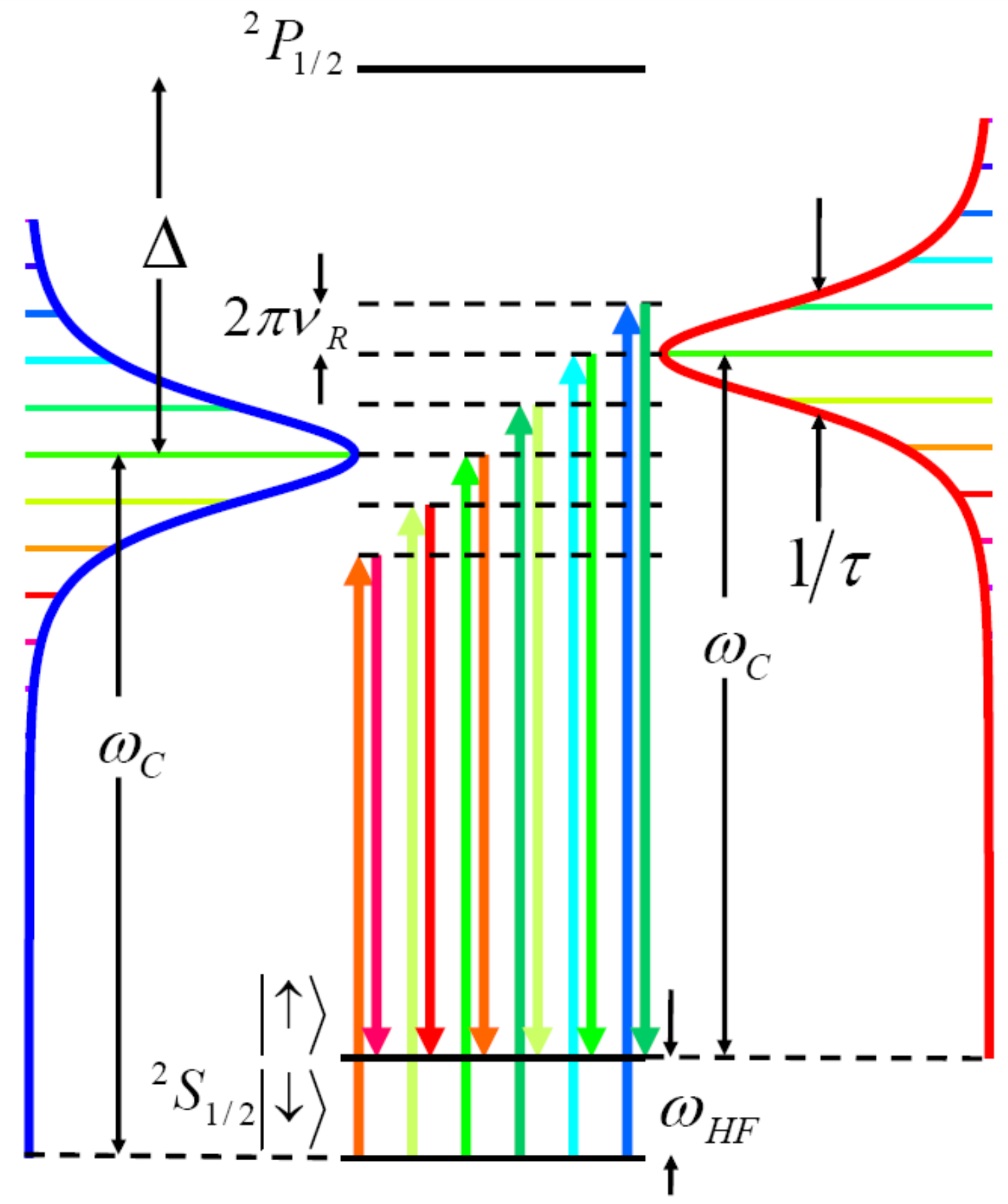}
\caption{The Stokes Raman process driven by frequency combs is shown here schematically.  An atom starting in the $\left|\downarrow\right\rangle$ state can be excited to a virtual level by absorbing a photon from the blue comb and then driven to the $\left|\uparrow\right\rangle$ state by emitting a photon into the red comb.  Although drawn here as two different combs, if the pulsed laser's repetition rate or one of its harmonics is in resonance with the hyperfine frequency, the absorption and emission can both be stimulated by the same frequency comb.  Because of the even spacing of the frequency comb, all of the comb teeth contribute through different virtual states which result in indistinguishable paths and add constructively.}
\label{fig:energydiagram}
\end{figure}

High fidelity qubit operations through Raman transitions are typically achieved by phase-locking frequency components separated by the energy difference of the qubit states.  
This is traditionally accomplished in a bottom-up type of approach where either two monochromatic lasers are phase-locked or a single cw laser is modulated by an acousto-optic (AO) or an electro-optic (EO) modulator.  However, the technical demands of phase-locked lasers and the limited bandwidths of the modulators hinder their application to experiments.  Here we exploit the large bandwidth of ultrafast laser pulses in a simple top-down approach toward bridging large frequency gaps and controlling complex atomic systems.  By starting with the broad bandwidth of an ultrafast laser pulse, a spectral landscape can be sculpted by interference from sequential pulses, pulse shaping and frequency shifting.  In this paper, we start with a picosecond pulse and, through the application of many pulses, generate a frequency comb that drives Raman transitions by stimulating absorption from one comb tooth and stimulating emission into another comb tooth as depicted in Fig. \ref{fig:energydiagram}.  Because this process only relies on the frequency difference between comb teeth, their absolute position is irrelevant and the carrier-envelope phase does not need to be locked \cite{cundiff:2003}.  As an example of how this new technique promises to ease experimental complexities, the control of metastable-state qubits separated by a terahertz was recently achieved using cw lasers that are phase-locked through a frequency comb \cite{urabe:2009}, but might be controlled \textit{directly} with a 100 fs Ti:sapph pulsed laser.

\begin{figure}[!]
\includegraphics[width=3.25in]{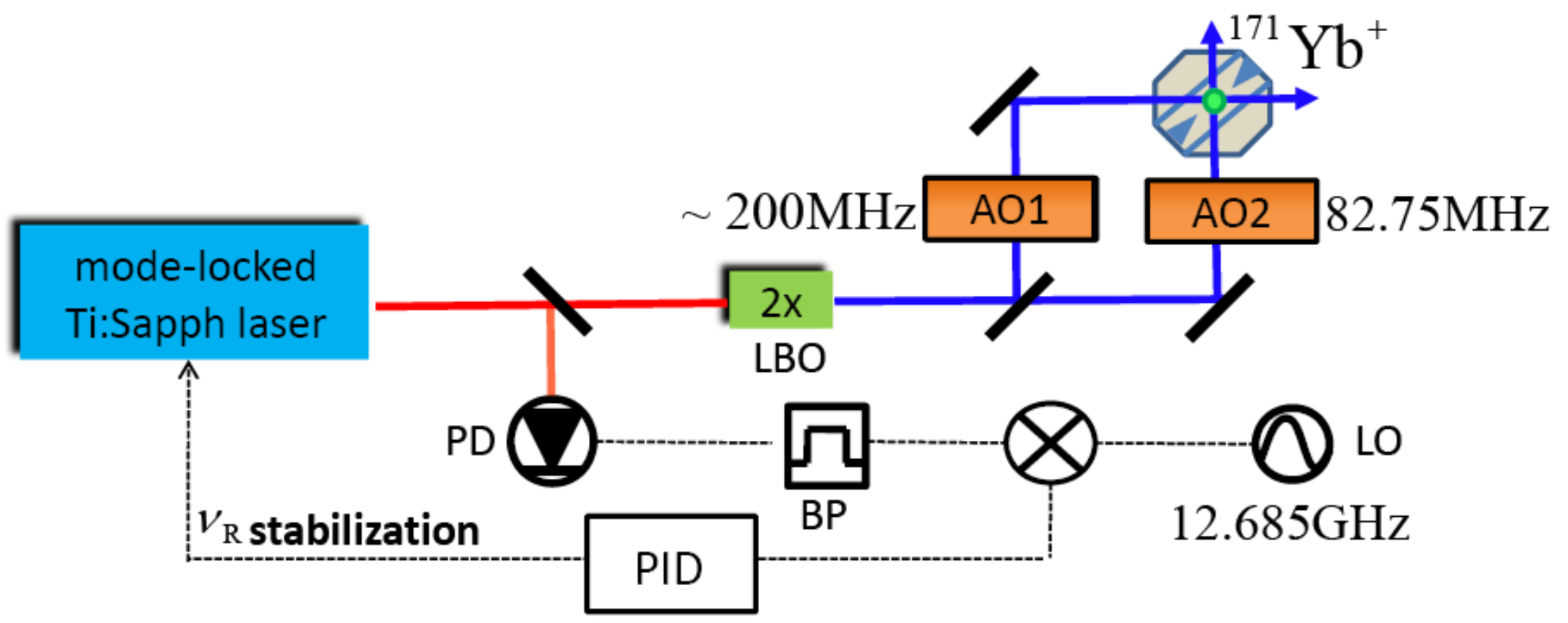}
\caption{
Schematic of the experimental setup showing the paths of the pulse trains emitted by a mode-locked Ti:Sapphire (Ti:Sapph) laser, where the optical pulses are frequency shifted by AOs. Single qubit rotations only require a single pulse train, but to address the motional modes the pulse train is split into two and sent through AOs to tune the relative offset of the two combs. We lock the repetition rate ($\nu_R$) by first detecting $\nu_R$ with a photodetector (PD). The output of the PD is an RF frequency comb spaced by $\nu_R$. 
We bandpass filter (BP) the RF comb at 12.685 GHz and then mix the signal with a local oscillator (LO). The
output of the mixer is sent into a feedback loop (PID) which stabilizes $\nu_R$ by means of a piezo mounted on one of the laser cavity mirrors.  When locked, $\nu_R$ is stable to within 1 Hz for more than an hour.  As an alternative, instead of locking the repetition rate of the pulsed laser, an error signal could be sent to one of the AOs to use the relative offset of the two combs to compensate for a change in the comb spacing.
}
\label{fig:experiment}
\end{figure}

\begin{figure}[b]
\includegraphics[width=3.0in]{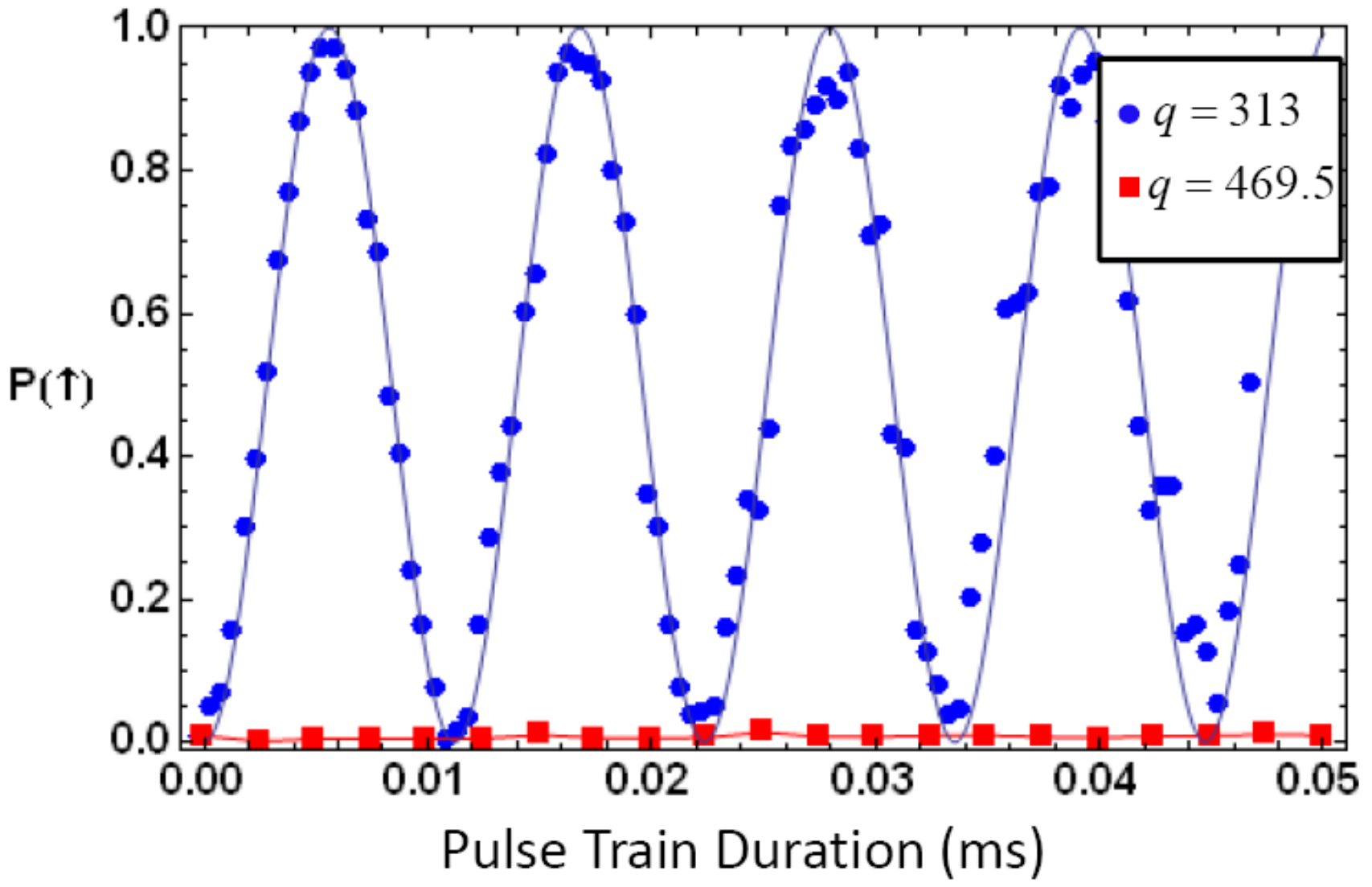}
\caption{After Doppler cooling and optical pumping to the $\left|\downarrow\right\rangle$ state, a single pulse train is directed onto the ion.  When the ratio of qubit splitting to pulse repetition rate, $q$, is an integer, pairs of comb teeth can drive Raman transitions as shown by the blue circular data points.  However, if the $q$ parameter is a half integer, the qubit remains in the initial state as shown by the red square data points.}
\label{fig:pulse:picking}
\end{figure}

At a fixed point in space, an idealized train of laser pulses has a time-dependent electric field that can be written as	   
\begin{equation}
E(t)=\sum^{N}_{n=1}f(t-nT)e^{i\omega_ct},
\label{eq:pulse:train}
\end{equation}
where $f(t)$ is the pulse envelope, $T$ is the time between successive pulses (repetition rate $\nu_{\mathrm{R}}=1/T$), $N$ is the number of pulses in the train and $\omega_c$ is the carrier frequency of the pulse.  For simplicity, any pulse-to-pulse optical phase shift is ignored since the offset frequency in the comb is unimportant.  The Fourier transform of Eq.~(\ref{eq:pulse:train}) defines a frequency comb characterized by an envelope $\tilde{f}(\omega)\equiv\mathscr{F}[f(t)]$ centered around the optical frequency $\omega_c$ and teeth separated by $\nu_R$ whose individual widths scale like $\sim\nu_R/N$.  The Raman resonance condition will be satisfied when a harmonic of the repetition rate is equal to the hyperfine qubit splitting $\omega_0$, implying that the parameter $q\equiv\omega_0/2\pi\nu_{R}$, is an integer.
To demonstrate coherent control with a pulse train, $^{171}\mathrm{Yb}^+$ ions confined in a linear Paul trap are used
to encode qubits in the $^2S_{1/2}$ hyperfine clock states $\left|{F=0, m_F = 0}\right\rangle \equiv \left|\downarrow\right\rangle$ and
$\left|{F=1, m_F = 0}\right\rangle \equiv \left|\uparrow\right\rangle$, having hyperfine splitting $\omega_0/2\pi = 12.6428$ GHz.
For state preparation and detection we use standard Doppler cooling, optical pumping, and state-dependent
fluorescence methods on the $811$ THz $^2S_{1/2} \leftrightarrow {^2P_{1/2}}$ electronic transition 
\cite{olmschenk:2007}.  The frequency comb is produced by a frequency-doubled mode-locked Ti:Sapphire laser at
a carrier frequency of $802$ THz, detuned by $\Delta/2\pi = 9$ THz from the electronic transition. 
The repetition rate of the laser is $\nu_R = 80.78$ MHz, with each pulse having a duration of 
$\tau \approx 1$ psec.  The repetition rate is phase-locked to a stable microwave oscillator as shown
in Fig. \ref{fig:experiment}, providing a ratio of hyperfine splitting to comb spacing of $q=156.5$.  An EO pulse picker
is used to allow the passage of one out of every $n$ pulses, decreasing the comb spacing by a factor of 
$n$ and permitting integral values of $q$.  As shown in Fig. \ref{fig:pulse:picking}, when $n=2$ ($q=313$ and $\nu_R = 40.39$ MHz),
application of the pulse train drives oscillations between the qubit states of a single ion.  However, when $n=3$ ($q=469.5$ and $\nu_R = 26.93$ MHz), the
qubit does not evolve.

The Rabi frequency of these oscillations can be estimated by considering the Hamiltonian resulting from an 
infinite train of pulses.  After adiabatically eliminating the excited $^2P_{1/2}$ state and performing the rotating-wave approximation, the resonant Rabi frequency of Raman transitions between the qubit states is 
given by a sum over all spectral components of the comb teeth as indicated in Fig. \ref{fig:energydiagram} ($\hbar =1$):
\begin{equation}
\Omega = \frac{|\mu|^2 \sum_l E_l E_{l-q}}{\Delta} 
       \approx \Omega_0 \left( \frac{\omega_0 \tau}{e^{\omega_0 \tau/2}-e^{-\omega_0 \tau/2}} \right),
\end{equation}
where $\mu$ is the dipole matrix element between the ground and excited electronic states,
$E_k \equiv \nu_R \tilde{f}(2\pi k \nu_R)$, and $q$ is an integer.  In the approximate expression 
above, the sum is replaced by an integral and each pulse is described by 
$f(t)=\sqrt{\pi/2}E_0 \mathrm{sech}(\pi t/\tau)$ with $\tau \ll T$, 
where $\Omega_0 = (\nu_R \tau) |\mu E_0|^2/\Delta = s\gamma^2/2\Delta$ is the time-averaged
resonant Rabi frequency of the pulse train and $s=\bar{I}/I_{sat}$ is the average intensity $\bar{I}=\nu_Rc\epsilon_0/2\int dt|f(t)|^2$
scaled to the $^2S_{1/2} \leftrightarrow {^2P_{1/2}}$ saturation intensity.  
Note the net transition rate is suppressed unless the single-pulse bandwidth is large compared to the
hyperfine frequency ($\omega_0 \tau \ll 1$), in which case $\Omega \approx \Omega_0$.  In our experiments, 
$\omega_0 \tau \approx 0.08$.  For $I_{sat} = 0.15$ W/cm$^2$, the data shown in Fig.~\ref{fig:pulse:picking} is consistent with an average intensity $\bar{I} \approx 500$ W/cm$^2$.

In order to entangle multiple ions, we first address the motion of the ion by resolving motional
sideband transitions.  As depicted in Fig. \ref{fig:experiment}, the pulse train is split into
two perpendicular beams with wavevector difference $k$ along the $x-$direction of motion.  Their 
polarizations are mutually orthogonal to each other and to a weak magnetic field that defines the quantization 
axis \cite{monroe:cooling}.  We control the spectral beatnotes between the combs by sending both  
beams through AO modulators (driven at frequencies $\nu_1$ and $\nu_2$), 
imparting a net offset frequency of $\Delta \omega/2\pi = \nu_1 - \nu_2$ between the combs.
For instance, in order to drive the first upper/lower sideband transition we set 
$|2\pi j \nu_R+\Delta \omega| = \omega_0\pm\omega_t$, with $j$ an integer and $\omega_t$
the trap frequency.  In order to see how the sidebands are spectrally resolved, we consider the following Hamiltonian of a single ion and single mode of harmonic motion interacting with the Raman pulse train:
\begin{widetext}
\begin{equation}
H_{\text{eff}} = \omega_t a^\dag{a} + \frac{\omega_0}{2}\sigma_z 
   + \frac{\theta_p}{2}\sum_n \delta (t-nT)
            \left( \sigma_+ e^{i(k\hat{x}+\Delta\omega t)} + \sigma_- e^{-i(k\hat{x}+\Delta\omega t)} \right) ,
\end{equation}
\end{widetext}
where $\theta_p = \Omega T$ is the change in the Bloch angle due to a single pulse, $\sigma_z$ is the Pauli-z operator,
$\sigma_\pm$ are raising and lowering operators, $\hat{x}$ is the $x-$position operator of the trapped ion, $a^\dag$ and $a$ are the raising and 
lowering operators of the $x-$mode of harmonic motion and the $q$ parameter has been assumed to not be an integer or half-integer. In the interaction picture, the evolution operator
after $N$ pulses is given by $V^{N}$, where

\begin{equation}
V = \text{exp}\left[-i H_0 T\right] \text{exp}\left[ \frac{-i\theta_p}{2}
                               \left(\sigma_+ e^{ik\hat{x}} + \sigma_- e^{-ik\hat{x}} \right) \right] 
\end{equation}
and $H_0 = \omega_t a^{\dag} a + 1/2(\omega_0 + \Delta\omega)\sigma_z$.
The time evolution operator is given by,
\begin{align}
\label{eq:FirstOrderV}
V^{N}&=\mathrm{exp}[-iH_0NT](\hat{I}-i\frac{\theta_p}{2}\sum_{n=0}^{N-1}Q_n+\mathcal{O}(\theta_p^2))\\
\label{eq:Qn}
Q_n&\equiv\sigma_+e^{i(\omega_0+\Delta\omega)nT}D(i\eta e^{i\omega_tnT})+h.c.~,
\end{align}
where $D(\alpha)=\mathrm{exp}[\alpha a^{\dagger}-\alpha^* a]$ is the harmonic oscillator displacement operator in phase space, and $\eta=k\sqrt{\hbar/2m\omega_t}$ is the Lamb-Dicke parameter.
In the Lamb-Dicke regime, $\eta\sqrt{\left\langle a^{\dagger}a\right\rangle+1}\ll1$, we can write $D(i\eta e^{i\omega_tnT})\approx1+i\eta(e^{i\omega_tnT}a^{\dagger}+e^{-i\omega_tnT}a)$ turning the sum in Eq.~(\ref{eq:FirstOrderV}) into a geometric series.  If, for example, the offset frequency between the combs $\Delta\omega$ is tuned to satisfy the resonance condition for the red sideband, $\vartheta_r\equiv(\omega_0+\Delta\omega-\omega_t)T=2\pi j$, where $j$ is an integer, then the sum in Eq. (\ref{eq:FirstOrderV}) is approximately given by, 
\begin{equation}
\sum_{n=0}^{N-1}Q_n\approx i\eta\frac{\mathrm{sin}N\vartheta_r/2}{\mathrm{sin}\vartheta_r/2}e^{i\vartheta_r(N-1)/2}\sigma_{+}a+h.c.
\label{eq:Red:SideBand}
\end{equation}
The coefficient in Eq.~(\ref{eq:Red:SideBand}) is the same as the field amplitude created by a diffraction grating of $N$ slits, whose narrow peaks have an amplitude equal to $N$.  In the limit $\omega_tT\ll1$, the other terms in Eq.~(\ref{eq:FirstOrderV}) that drive the carrier and other sideband transitions can be neglected when $N\gg(\omega_tT\eta)^{-1}$.  This is analogous to the destructive interference of amplitudes away from the bright peaks in a diffraction grating.  For $\omega_t/2\pi=1.64~\mathrm{MHz}$, $T=12.4~\mathrm{ns}$ and $\eta=0.1$, the sidebands are well-resolved when $N\gg80$.
\begin{figure}[!]
\includegraphics[width=3.0in]{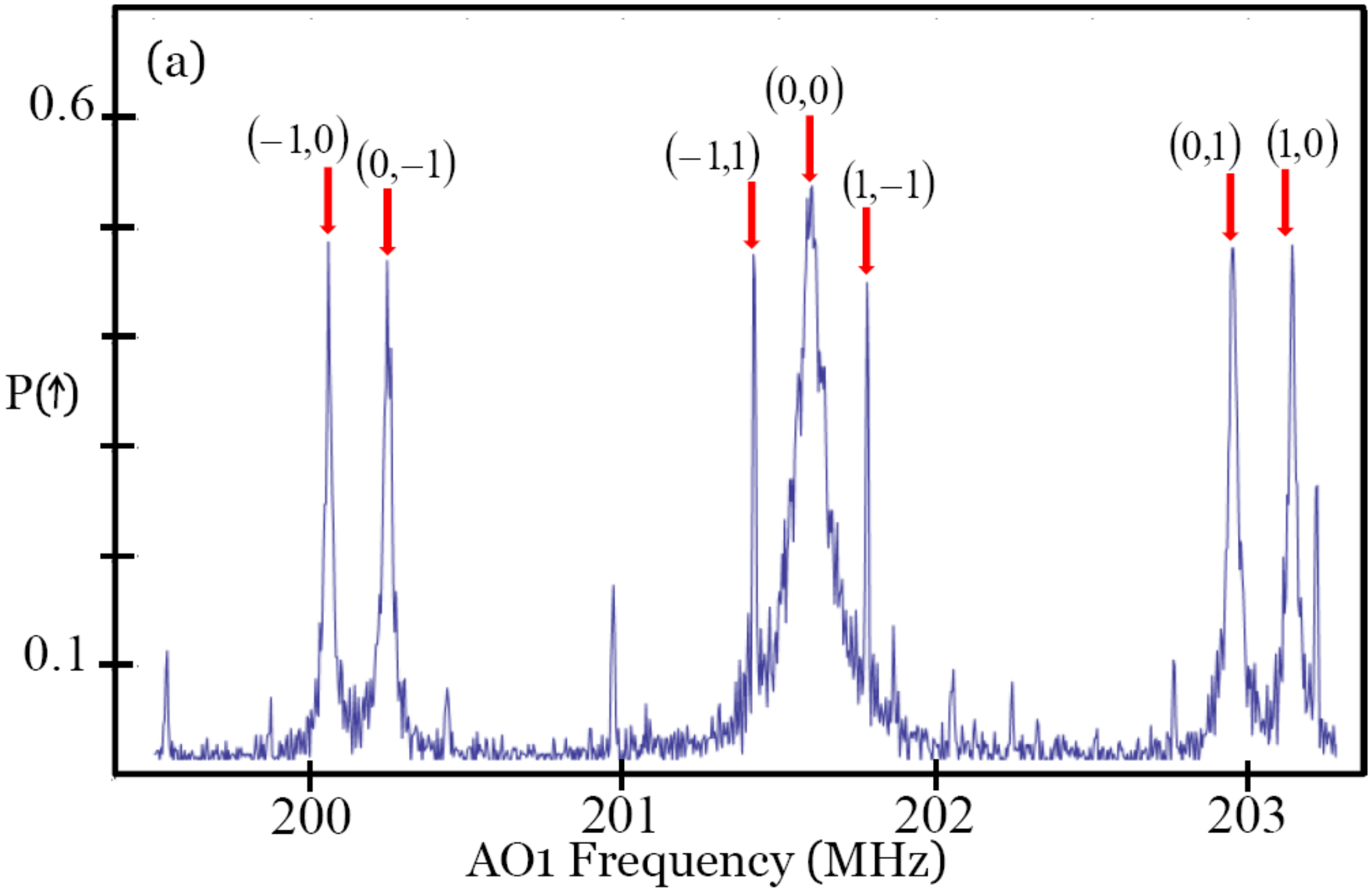}
\includegraphics[width=3.25in]{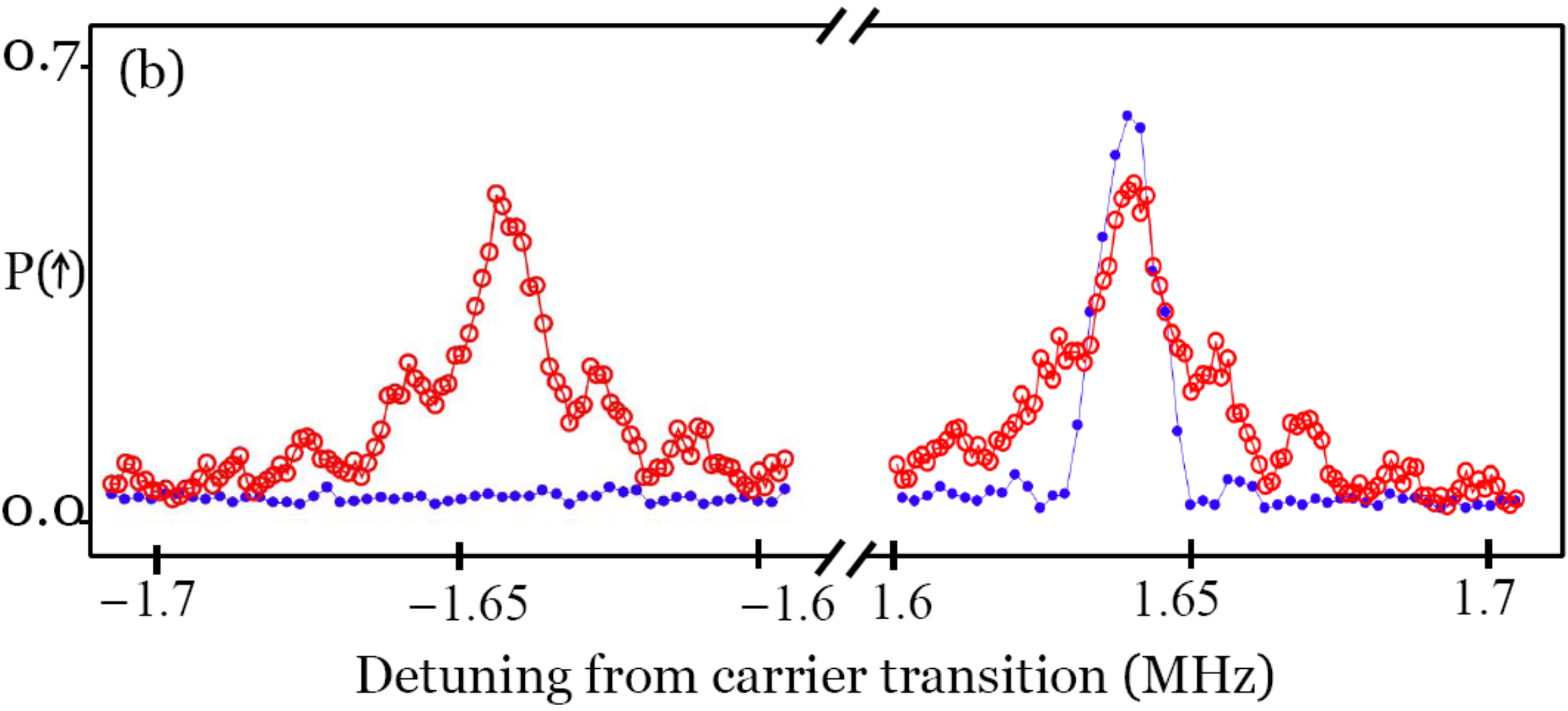}
\caption{(a) Using a Raman probe duration of $80\mu s$, ($N\sim6500$), a frequency scan of AO1 shows the resolved carrier and motional sideband transitions of a single trapped ion.  The transitions are labeled, $(\Delta n_x,\Delta n_y)$, to indicate the change in the number of phonons in the two transverse modes that accompany a spin flip. The x and y mode splitting is controlled by applying biasing voltages to the trap electrodes.  Unlabeled peaks show higher order sideband transitions and transitions to other Zeeman levels due to imperfect polarization of the Raman beams. (b) Ground state cooling of the motional modes via a train of phase-coherent ultra-fast pulses.  The red open-circle data points show that after Doppler cooling and optical pumping, both the red and blue sidebands are easily driven.  The blue filled-circle data points show that after sideband cooling, the ion is close to the motional ground state, ($\bar{n}_{x,y}\leq0.03$), as evidenced by the suppression of the red-sideband transition.}
\label{fig:cooling}
\end{figure}

For many applications in quantum information, the motional modes of the ion must be cooled and initialized to a nearly pure state.  Fig.~\ref{fig:cooling} shows that the pulsed laser can also be used to carry out the standard techniques of sideband cooling \cite{monroe:cooling} to prepare the ion in the motional ground state with near unit fidelity.  The set-up also easily lends itself to implementing a two-qubit entangling gate by applying two fields whose frequencies are symmetrically detuned from the red and blue sidebands \cite{molmer:sorensen:1999,james:2000}.  By simultaneously applying two modulation frequencies to one of the comb AO frequency shifters, we create two combs in one of the beams.  When these combs are tuned to drive the red and blue sidebands (in conjunction with the third frequency comb in the other beam), the ion experiences a spin-dependent force in a rotated basis as described in Ref. \cite{haljan:2005}.  Ideally, when the fields are detuned from the sidebands by an equal and opposite amount  $\delta = 2\eta\Omega$, a decoupling of the motion and spin occurs at gate time $t_g = 2\pi/\delta,$ and the spin state evolve to the maximally entangled state $|\chi\rangle=\left|\downarrow\downarrow\right\rangle+e^{i\varphi}\left|\uparrow\uparrow\right\rangle$.  In the experiment, $t_g = 108 \mu$s ($N \sim 8700$ pulses).
  
\begin{figure}[!]
\includegraphics[width=2.8in]{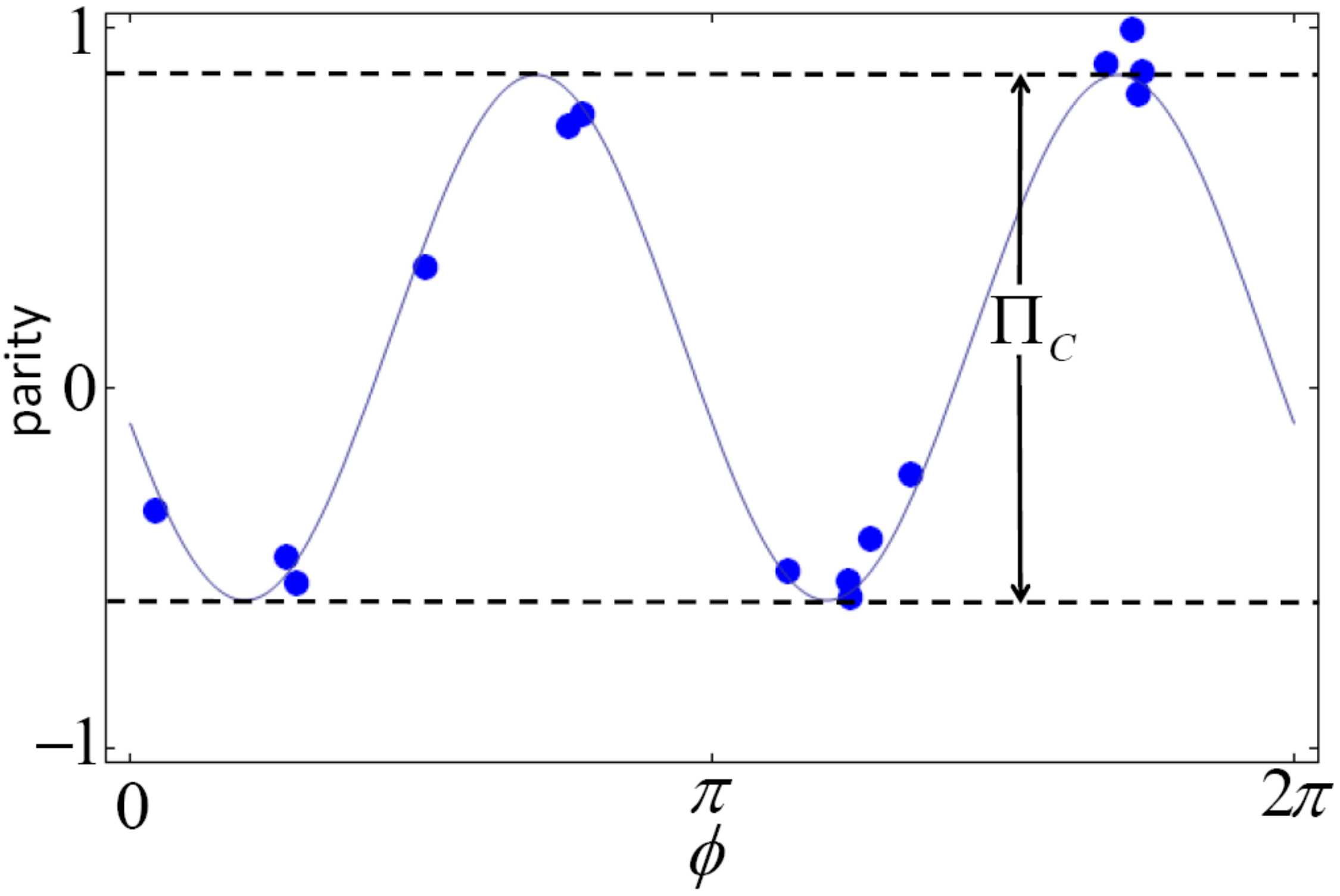}
\caption{The parity oscillation that is used to calculate the fidelity of the spin state of two ions with respect to the maximally entangled state $|\chi\rangle$ after performing the entangling gate.  The phase $\phi$ of the analyzing pulse is scanned by changing the relative phase of the rotation pulses.  The offset and lack of full contrast in the parity signal can be attributed to state detection errors.}
\label{fig:TwoIonGate}
\end{figure}

The entanglement is verified by the measurement of a fidelity-based entanglement witness \cite{bourennane:2004}.  The fidelity of the state $\rho$ with respect to $\left|\chi\right\rangle$ is found by measuring the populations $\rho_{\uparrow\uparrow,\uparrow\uparrow}$ and $\rho_{\downarrow\downarrow,\downarrow\downarrow}$ and scanning the measurement phase angle $\phi$ in the parity signal $\Pi(\phi)=Tr[\sigma^{(1)}_{z}\sigma^{(2)}_{z} {R(\phi)}^{\otimes 2}\rho {R^{\dagger}(\phi)}^{\otimes 2}]$, where $R(\phi)$ is a $\pi/2$ rotation on the Bloch sphere with phase $\phi$ \cite{sackett:2000}.  The contrast of the parity signal, $\Pi_C$, is used to calculate the fidelity $\mathcal{F}=(\rho_{\uparrow\uparrow,\uparrow\uparrow}+\rho_{\downarrow\downarrow,\downarrow\downarrow})/2+\Pi_C/4$.  The measured populations of $\left|\downarrow\downarrow\right\rangle$ and $\left|\uparrow\uparrow\right\rangle$ together with the data shown in Fig. \ref{fig:TwoIonGate} yield a fidelity $\mathcal{F}=0.86\pm0.03$, thereby signaling that the two ions are entangled after the application of the pulse train.

We have demonstrated full control and entanglement of two atomic qubits using an optical frequency comb.  This work represents a significant simplification over current methods for optical control of trapped ion qubits, and also points the way toward future advances with higher power laser pulses.  For example, such pulses allow much larger detunings from resonance and a suppression of decoherence from spontaneous emission \cite{wineland:2005} while maintaining gate speed.  When only a few high-power pulses are used in ways similar to the experiment, it also becomes possible to suppress sources of motional decoherence through the use of fast entanglement schemes \cite{garcia:2003}.

\begin{acknowledgments}
This work is supported by the Army Research Office (ARO) with funds 
from the DARPA Optical Lattice Emulator (OLE) Program, IARPA under 
ARO contract, the NSF Physics at the Information Frontier Program, 
and the NSF Physics Frontier Center at JQI.
\end{acknowledgments}
%

%
%

\end{document}